\documentclass[showpacs,preprintnumbers,amsmath,amssymb,nofootinbib]{revtex4}

\usepackage{setspace}

\usepackage{graphicx}  % Include figure files
\usepackage{dcolumn}   % Align table columns on decimal point
\usepackage{bm}        % bold math

\newcommand{\ETM}       {E_T\hspace{-2.4ex}/\hspace{1.2ex}}

\begin{document}

%\preprint{hep-ex/05xxxix}
%\date{\today}

% ======================
% TITLE AND ABSTRACT
% ======================

\begin{center}
\begin{large}
\bf{Study of Jet Shapes \\ in Inclusive Jet Production in
$p \bar p$ Collisions at $\sqrt{s} = {\rm 1.96 \ TeV}$ }
\end{large}
\end{center}

\font\eightit=cmti8
\def\r#1{\ignorespaces $^{#1}$}

\hfilneg
\begin{sloppypar}
\noindent 
D.~Acosta,\r {16} J.~Adelman,\r {12} T.~Affolder,\r 9 T.~Akimoto,\r {54}
M.G.~Albrow,\r {15} D.~Ambrose,\r {15} S.~Amerio,\r {42}
D.~Amidei,\r {33} A.~Anastassov,\r {50} K.~Anikeev,\r {15}
% ...
\end{sloppypar}
\vskip .026in
\begin{center}
(CDF Collaboration)
\end{center}

\vskip .026in
\begin{center}
\r 1  {\eightit Institute of Physics, Academia Sinica, Taipei, Taiwan 11529,
Republic of China} \\
\r 2  {\eightit Argonne National Laboratory, Argonne, Illinois 60439} \\
\r 3  {\eightit Institut de Fisica d'Altes Energies, Universitat Autonoma
de Barcelona, E-08193, Bellaterra (Barcelona), Spain} \\
\r 4  {\eightit Istituto Nazionale di Fisica Nucleare, University of Bologna,
I-40127 Bologna, Italy} \\
\r 5  {\eightit Brandeis University, Waltham, Massachusetts 02254} \\
\r 6  {\eightit University of California, Davis, Davis, California  95616} \\
\r 7  {\eightit University of California, Los Angeles, Los
Angeles, California  90024} \\
\r 8  {\eightit University of California, San Diego, La Jolla, California  92093} \\
\r 9  {\eightit University of California, Santa Barbara, Santa Barbara, California
93106} \\
\r {10} {\eightit Instituto de Fisica de Cantabria, CSIC-University of Cantabria,
39005 Santander, Spain} \\
\r {11} {\eightit Carnegie Mellon University, Pittsburgh, PA  15213} \\
\r {12} {\eightit Enrico Fermi Institute, University of Chicago, Chicago,
Illinois 60637} \\
\r {13}  {\eightit Joint Institute for Nuclear Research, RU-141980 Dubna, Russia}
\\
\r {14} {\eightit Duke University, Durham, North Carolina  27708} \\
\r {15} {\eightit Fermi National Accelerator Laboratory, Batavia, Illinois
60510} \\
\r {16} {\eightit University of Florida, Gainesville, Florida  32611} \\
\r {17} {\eightit Laboratori Nazionali di Frascati, Istituto Nazionale di Fisica
               Nucleare, I-00044 Frascati, Italy} \\
\r {18} {\eightit University of Geneva, CH-1211 Geneva 4, Switzerland} \\
\r {19} {\eightit Glasgow University, Glasgow G12 8QQ, United Kingdom}\\
\r {20} {\eightit Harvard University, Cambridge, Massachusetts 02138} \\
\r {21} {\eightit Division of High Energy Physics, Department of
Physics, University of Helsinki and Helsinki Institute of Physics,
FIN-00014, Helsinki, Finland}\\
\r {22} {\eightit Hiroshima University, Higashi-Hiroshima 724, Japan} \\
\r {23} {\eightit University of Illinois, Urbana, Illinois 61801} \\
\r {24} {\eightit The Johns Hopkins University, Baltimore, Maryland 21218} \\
\r {25} {\eightit Institut f\"{u}r Experimentelle Kernphysik,
Universit\"{a}t Karlsruhe, 76128 Karlsruhe, Germany} \\
\r {26} {\eightit High Energy Accelerator Research Organization (KEK), Tsukuba,
Ibaraki 305, Japan} \\
\r {27} {\eightit Center for High Energy Physics: Kyungpook National
University, Taegu 702-701; Seoul National University, Seoul 151-742; and
SungKyunKwan University, Suwon 440-746; Korea} \\
\r {28} {\eightit Ernest Orlando Lawrence Berkeley National Laboratory,
Berkeley, California 94720} \\
\r {29} {\eightit University of Liverpool, Liverpool L69 7ZE, United Kingdom} \\
\r {30} {\eightit University College London, London WC1E 6BT, United Kingdom} \\
\r {31} {\eightit Massachusetts Institute of Technology, Cambridge,
Massachusetts  02139} \\
\r {32} {\eightit Institute of Particle Physics: McGill University,
Montr\'{e}al, Canada H3A~2T8; and University of Toronto, Toronto, Canada
M5S~1A7} \\
\r {33} {\eightit University of Michigan, Ann Arbor, Michigan 48109} \\
\r {34} {\eightit Michigan State University, East Lansing, Michigan  48824} \\
\r {35} {\eightit Institution for Theoretical and Experimental Physics, ITEP,
Moscow 117259, Russia} \\
\r {36} {\eightit University of New Mexico, Albuquerque, New Mexico 87131} \\
\r {37} {\eightit Northwestern University, Evanston, Illinois  60208} \\
\r {38} {\eightit The Ohio State University, Columbus, Ohio  43210} \\
\r {39} {\eightit Okayama University, Okayama 700-8530, Japan}\\
\r {40} {\eightit Osaka City University, Osaka 588, Japan} \\
\r {41} {\eightit University of Oxford, Oxford OX1 3RH, United Kingdom} \\
\r {42} {\eightit University of Padova, Istituto Nazionale di Fisica
          Nucleare, Sezione di Padova-Trento, I-35131 Padova, Italy} \\
\r {43} {\eightit University of Pennsylvania, Philadelphia,
        Pennsylvania 19104} \\
\r {44} {\eightit Istituto Nazionale di Fisica Nucleare Pisa, Universities 
of Pisa, Siena and Scuola Normale Superiore, I-56127 Pisa, Italy} \\
\r {45} {\eightit University of Pittsburgh, Pittsburgh, Pennsylvania 15260} \\
\r {46} {\eightit Purdue University, West Lafayette, Indiana 47907} \\
\r {47} {\eightit University of Rochester, Rochester, New York 14627} \\
\r {48} {\eightit The Rockefeller University, New York, New York 10021} \\
\r {49} {\eightit Istituto Nazionale di Fisica Nucleare, Sezione di Roma 1,
University di Roma ``La Sapienza," I-00185 Roma, Italy}\\
\r {50} {\eightit Rutgers University, Piscataway, New Jersey 08855} \\
\r {51} {\eightit Texas A\&M University, College Station, Texas 77843} \\
\r {52} {\eightit Texas Tech University, Lubbock, Texas 79409} \\
\r {53} {\eightit Istituto Nazionale di Fisica Nucleare, University of Trieste/\
Udine, Italy} \\
\r {54} {\eightit University of Tsukuba, Tsukuba, Ibaraki 305, Japan} \\
\r {55} {\eightit Tufts University, Medford, Massachusetts 02155} \\
\r {56} {\eightit Waseda University, Tokyo 169, Japan} \\
\r {57} {\eightit Wayne State University, Detroit, Michigan  48201} \\
\r {58} {\eightit University of Wisconsin, Madison, Wisconsin 53706} \\
\r {59} {\eightit Yale University, New Haven, Connecticut 06520} \\
\end{center}

\begin{abstract}

We report on a study of jet shapes 
in inclusive jet production in $p \overline{p}$ collisions at $\sqrt{s} = 1.96 \ {\rm TeV}$ 
using the upgraded Collider Detector at Fermilab in Run II (CDF II) and based on an integrated luminosity 
of $170 \ \rm pb^{-1}$. Measurements are carried out on jets with rapidity $0.1 < |Y^{\rm jet}| < 0.7$ 
and transverse momentum $37$ GeV/c $< P_T^{\rm jet} < 380$~GeV/c. The jets have been corrected to the hadron level.
The measured jet shapes are compared to leading-order QCD parton-shower Monte Carlo predictions as 
implemented in the PYTHIA and HERWIG programs. 
PYTHIA, tuned to describe the underlying event as measured in CDF Run I,
provides a better description of the measured jet shapes than does
PYTHIA or HERWIG with their default parameters.

\end{abstract}

%\keywords{Suggested keywords} % Use showkeys class option if keyword
                               % display desired

\maketitle

% ==================================
% INTRODUCTION
% ==================================

\section{Introduction}

The measurement of the jet shape allows a  study of the transition between a parton produced in 
a hard process and the collimated flow of hadrons observed experimentally~\cite{jetshape}. 
The internal structure of a jet is dominated by multi-gluon emissions from the primary outgoing parton and 
is expected to depend mainly on the type of parton, quark or gluon, creating the jet and the transverse momentum of the jet.
In hadron-hadron collisions, the jet shape also receives contributions from initial-state radiation emitted from the colliding partons 
and multiple parton interactions between remnants (the so-called {\it{underlying event}}). The effects of initial-state radiation 
are described by the parton showering in QCD Monte Carlo programs while the underlying event description is provided by phenomenological models.
The comparison of jet cross section measurements  with perturbative QCD predictions, as well as
the estimation of QCD backgrounds in the search for new physics, requires an accurate description of
the underlying event. The study of jet shapes at the Tevatron provides a precise means to test 
the validity of the models for parton cascades and the underlying event in hadron-hadron collisions. 
Measurements of the jet shape have been performed in $p \overline{p}$ collisions at   
$\sqrt{s} = 1.8 \ {\rm TeV}$~\cite{pp}, deeply inelastic scattering 
(DIS)~\cite{ep} and photoproduction \cite{gp} processes in $e^{\pm}p$ collisions at 
HERA, and  $e^{+}e^{-}$ interactions at LEP1~\cite{ee}. It was observed~\cite{ee} that the jets in $p \overline{p}$ collisions
are significantly broader than those in $e^{+}e^{-}$  with most of the difference being explained in terms of the different mixtures of 
quark and gluon jets in the final state. The jets in DIS were found to be very similar to those in  $e^{+}e^{-}$ interactions
and narrower than those in $p \overline{p}$ collisions. In this paper, new jet shape results  in $p \overline{p}$ collisions,  
based on CDF Run II data, are presented for central jets in a wide range of jet transverse momentum. For the first time, these measurements  extend the study of jet internal structure to jets with transverse momentum up to $380 \ {\rm GeV/c}$.

% ==================================
% DETECTOR SETUP
% ==================================

\section{Experimental setup}

\begin{figure}[tbh]
\centerline{\includegraphics{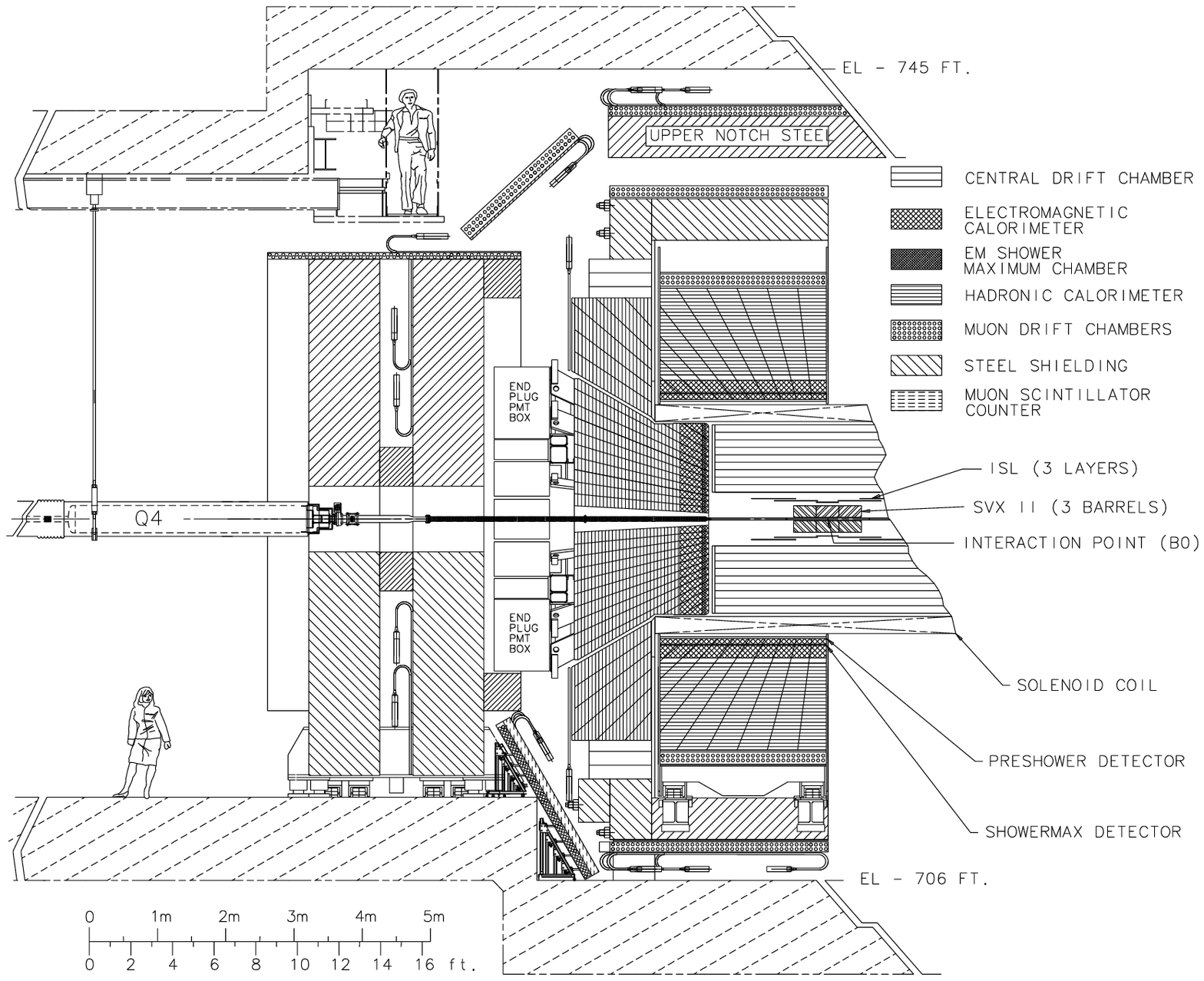}} 
\caption{Longitudinal view of half of the CDF II detector.} 
\label{fig1}
\end{figure}

The CDF II detector is described in detail in \cite{cdfii}. 
In this section, the sub-detectors most relevant 
for this analysis are briefly discussed. As illustrated in Fig.~\ref{fig1}, the detector has a 
charged particle tracking system immersed in a 1.4~T magnetic field, aligned coaxially with the beam line.
A silicon microstrip detector~\cite{silicon} provides tracking over the radial range 1.35 to 28 cm. A 3.1 m long open-cell
drift chamber, the Central Outer Tracker (COT)~\cite{cot}, covers the radial range from 44 to 132 cm. The fiducial region
of the silicon detector covers the pseudorapidity~\cite{rapidity} range $|\eta| \leq 2$, while the COT provides 
coverage for $|\eta| \leq 1$. The charged particles are reconstructed in the COT with  a transverse-momentum resolution   
of $\sigma(p_T)/p^2_T \sim 1.7 \cdot 10^{-3} [\rm GeV/c]^{-1}$. 
Segmented sampling calorimeters, arranged in a projective tower
geometry, surround the tracking system and measure the energy flow of interacting particles in 
$|\eta| \leq 3.6$. The CDF central barrel calorimeter~\cite{ccal} is unchanged from Run I and 
covers the region $|\eta| < 1$. It consists of an electromagnetic (CEM) calorimeter and an hadronic (CHA) calorimeter 
segmented into  480 towers of size $0.1$ in $\eta$ and  $15^o$ in $\phi$.  
The end-wall hadronic (WHA) calorimeter~\cite{wha} complements the coverage of the central barrel calorimeter in the region 
$0.6 < |\eta| < 1.0$ and provides additional forward coverage out to $|\eta| < 1.3$.  
In Run II, new forward scintillator-plate calorimeters~\cite{pcal} replaced the original Run I gas calorimeter system.
The new plug electromagnetic (PEM) calorimeter covers the region $1.1 < |\eta| < 3.6$ while the new hadronic (PHA) calorimeter
provides coverage in the $1.3 < |\eta| < 3.6$ region. Each plug calorimeter is segmented into 480 towers with sizes that 
vary as a function of $\eta$ (0.1 in $\eta$ and $7.5^o$ in $\phi$ for $|\eta| < 1.8$  to 0.6 in $\eta$ and  $15^o$ in 
$\phi$ at $|\eta| = 3.6$).  The calorimetry    
has a crack at $\eta = 0$ (between the two halves of the central barrel calorimeter) and two cracks at $\eta = \pm 1.1$
(in the region between the WHA and the plug calorimeters). The measured energy resolutions for electrons in the electromagnetic
calorimeters are $14 \% /\sqrt{E_T}$ (CEM) and  $16 \% /\sqrt{E} \oplus 1 \%$ (PEM) where the units are expressed in GeV.
The single-pion energy resolutions in the hadronic calorimeters, as determined with 
test-beam data, are $75 \% /\sqrt{E_T}$ (CHA), $80 \% /\sqrt{E}$ (WHA) and $80 \% /\sqrt{E} \oplus 5 \%$ (PHA).
Cherenkov counters located in the $3.7 < |\eta| < 4.7$ region \cite{clc} measure the average 
number of inelastic $p \overline{p}$ collisions per bunch crossing and thereby determine the beam luminosity.
Finally, a three-level trigger system~\cite{trigger} is used to select events online, as described in section V.

% ========================
% MC SIMULATION
% ========================

\section{Monte Carlo simulation}
Monte Carlo event samples are used to determine the response of the detector and the 
correction factors to the hadron level~\cite{hadron} for the measured jet shapes. The generated samples are passed through a full 
CDF detector simulation (based on  GEANT3~\cite{geant} where the GFLASH~\cite{gflash} 
package is used to simulate the energy deposition in the calorimeters), and then reconstructed and analyzed using the same analysis chain as 
in the data. Samples of simulated inclusive jet events have been generated using
the PYTHIA 6.203~\cite{pythia} and HERWIG 6.4~\cite{herwig} Monte Carlo generators.
In both programs, the partonic interactions are generated
using leading-order QCD matrix elements, including initial- and final-state parton showers. 
CTEQ5L~\cite{cteq} parton distribution functions are used for the proton and antiproton. 
The HERWIG samples have been generated using default parameters.
The PYTHIA samples have been created using a special tuned set of parameters, denoted as 
PYTHIA-Tune A~\cite{tunea},  that includes  enhanced contributions from initial-state
gluon radiation and  secondary parton interactions between remnants. Tune A  
was determined as a result of dedicated studies of the underlying event in dijet events performed 
using the CDF Run I data \cite{underlying}.  
In addition, two different PYTHIA samples have been generated using the default parameters with and without   
the contribution from multiple parton interactions (MPI) between the proton and antiproton remnants. The latter 
are denoted as PYTHIA-(no MPI). The  HERWIG samples do not include multiple parton interactions.
Fragmentation 
into hadrons is carried out using the string model \cite{string} as implemented in JETSET~\cite{jetset}
in the case of PYTHIA and the cluster model~\cite{cluster} in HERWIG.

% ========================
% JET RECONSTRUCTION AND CORRECTION
% ========================
\section{Jet reconstruction}

An iterative cone-based midpoint algorithm \cite{midpoint} in the $Y$-$\phi$ plane~\cite{rapidity}  
is used to reconstruct jets from the energy deposits in the calorimeter towers for 
both data and the Monte Carlo simulated events, and 
from final-state particles  for the Monte Carlo generated events. This procedure is explained 
in detail below for the jet reconstruction from the 
calorimeter towers. In the first step, the electromagnetic and hadronic sections of each calorimeter tower 
are preclustered into a {\it{physics}} tower. The position of each section is determined from 
the unit vector joining the vertex of the interaction and the section's geometrical center.  
Each section  is assumed to be massless. 
The four-vector components  of each {\it{physics}} tower are then computed using the four-momentum sum of its electromagnetic 
and hadronic sections; only towers with transverse momentum above 0.1~GeV/c are further considered.  In a second step,  each 
{\it{physics}} tower 
with transverse momentum above 1~GeV/c is used to define a seed for the jet search.  Starting from the seed with highest transverse 
momentum, a  cone is drawn around each seed and
the {\it{physics}} towers inside a distance $\sqrt{(\Delta Y)^2 + (\Delta \phi)^2} < R/2$, with $R=0.7$ are used to determine 
the direction of the new cluster as indicated in Eqs.~\ref{eq1} and~\ref{eq2}:     
\begin{equation}
E^{\rm cluster} = \sum_{\rm phys. towers} E^{\rm tower}, \ \  P_{i}^{\rm cluster} = \sum_{\rm phys. towers} P_{i}^{tower} \ \ i=x,y,z 
\label{eq1}
\end{equation}

\begin{equation}
P_{T}^{\rm cluster} = \sqrt{(P_{x}^{\rm cluster})^2 + (P_{y}^{\rm cluster})^2}, \ \ 
Y^{\rm cluster} = \frac{1}{2} \frac{E^{\rm cluster}+P_{z}^{\rm cluster}}{E^{\rm cluster}-P_{z}^{\rm cluster}}, \ \
\phi^{\rm cluster} = tan^{-1}(\frac{P_{y}^{\rm cluster}}{P_{x}^{\rm cluster}})
\label{eq2}
\end{equation}
\noindent
where $Y^{\rm cluster}$ and $\phi^{\rm cluster}$ denote the rapidity and azimuthal angle of the cluster, respectively. Starting from the list of resulting clusters, the procedure is iterated until the contents of the clusters remain unchanged.
In a third step, the midpoint ($Y$-$\phi$ plane) between each pair of stable clusters separated by less than $2R$ is added to the list of clusters. The clustering algorithm, as explained above, is again iterated until stability is achieved. This latter 
step gives the name to the jet algorithm and was introduced in order to address the theoretical 
difficulties~\cite{seymour} of the cone-based jet algorithm used in Run I~\cite{jetclu}.
Finally, the cone size is expanded from $R/2$ to $R$~\cite{midpoint} and 
the momentum  sharing of overlapping clusters is considered. Overlapping jets are merged if their 
shared momentum is larger than $75 \%$  of the jet with smaller transverse 
momentum; otherwise two jets are formed and the common towers are assigned to the nearest jet. 
The variables for jets reconstructed from the calorimeter towers are denoted by $P_{T, \rm CAL}^{\rm jet}$, $Y^{\rm jet}_{\rm CAL}$ and $\phi^{\rm jet}_{\rm CAL}$. As mentioned, the same jet algorithm is applied to the final-state hadrons in Monte Carlo generated events. In this case, the four-vector components of each individual hadron are used as input to the algorithm  and no cut on the minimum transverse momentum of the particles is applied. The variables of the hadron-level jets are denoted by 
$P_{T, \rm HAD}^{\rm jet}$, $Y^{\rm jet}_{\rm HAD}$ and $\phi^{\rm jet}_{\rm HAD}$. \\

The reconstruction of the jet variables in the calorimeter  
is studied using Monte Carlo event samples and matched pair of jets at the calorimeter and hadron levels.  These studies
indicate that the angular variables of the jet,  $Y^{\rm jet}_{\rm CAL}$ and $\phi^{\rm jet}_{\rm CAL}$,  are
reconstructed in the calorimeter with no significant systematic shift and with a resolution, for jets with $P_{T, \rm CAL}^{\rm jet} > 20 \ {\rm GeV/c}$, 
of the order of 0.02 units and 0.025 units, respectively. The resolutions improve as the measured jet transverse momentum increases. 
The jet transverse momentum measured in the calorimeter, $P_{T, \rm CAL}^{\rm jet}$, systematically underestimates that of the 
hadron level jet. This is mainly due to the non-compensating nature of the calorimeter~\cite{calor}. For jets with  
$P_{T, \rm CAL}^{\rm jet} > 20 \ {\rm GeV/c}$ the jet transverse momentum is reconstructed with 
an average shift of $-20 \% $ and an r.m.s of 
$ 17 \% $. The reconstruction of the jet transverse momentum improves as $P_{T, \rm CAL}^{\rm jet}$ increases. For jets with   
$P_{T, \rm CAL}^{\rm jet} > 130 \ {\rm GeV/c}$ the jet transverse momentum is reconstructed with 
an average shift of $-12 \% $ and an r.m.s of 
$ 9 \% $. An average correction is extracted from the Monte Carlo 
using the following procedure: matched pairs of jets are used to study the difference between the jet transverse momentum at 
the hadron level, $P_{T, \rm HAD}^{\rm jet}$, and  the corresponding measurement in the calorimeter, $P_{T, \rm CAL}^{\rm jet}$. The 
resulting correlation is used to extract multiplicative correction factors, $C(P_{T,\rm CAL}^{\rm jet})$, 
which are then applied to the measured jets to obtain the corrected jet transverse momenta, $P_{T, \rm COR}^{\rm jet} = C \times P_{T,\rm CAL}^{\rm jet}$~\cite{olga}.

% ========================
% EVENT SELECTION
% ========================
\section{Event selection}
This analysis is based on a  
sample of inclusive jet events selected from the CDF Run II data corresponding to
a total integrated luminosity of 
170 ${\rm pb}^{-1}$. Events were collected {\it{online}} using three-level 
trigger paths, based on the measured energy deposits in the calorimeter towers,  
with several different thresholds on the jet transverse energies. In the first-level trigger, a single trigger tower 
with transverse energy above 5 GeV or 10 GeV, depending on the trigger path, is required.  
In the second-level trigger, a hardware-based clustering is carried out  where calorimeter clusters are 
formed around the selected trigger towers. The events are required to have at least one second-level trigger cluster 
with transverse energy above a given threshold, which varies between 15 and 90 GeV for the different trigger paths. In the 
third-level trigger,  jets are reconstructed using the CDF Run I cone algorithm~\cite{jetclu} and the events 
are required to have at least one jet with transverse energy above 20 to 100 GeV depending on the trigger path.   
Offline, jets are reconstructed using the midpoint algorithm, as explained above, starting from seed calorimeter 
towers with transverse momentum above 1 GeV/c and only considering towers with a 
minimum transverse momentum of 100 MeV/c in the clustering procedure. 
The following selection criteria have
 been imposed:
\begin{itemize}
\item One reconstructed primary vertex  with $z$-component, $V_Z$, in the region 
$|V_Z| < 60 {\rm \ cm}$. Events with more than one primary vertex are removed to eliminate contributions from pile-up events with 
multiple proton-antiproton interactions per beam crossing. The data used in this study was collected at Tevatron instantaneous 
luminosities in the range between $0.2 \times 10^{31} {\rm cm^{-2} s^{-1}}$ and  $4 \times 10^{31} {\rm cm^{-2} s^{-1}}$ for which, on average,  
less than one interaction per crossing is expected.  

\item $ \ETM/\sqrt{E_T} < 3.5 \ {\rm GeV}^{1/2}$, where $\ETM$ ($E_T$) denotes the missing (total) 
transverse energy of the event as determined from the energy deposits in the calorimeter towers. This cut eliminates  beam-related
backgrounds, beam halo and beam-gas contributions, and cosmic rays.

\item At least one jet with $P_{T \rm  ,COR}^{\rm jet} >  37$ GeV/c 
and $Y^{\rm jet}$ in the region $0.1 < |Y^{\rm jet}| < 0.7$. 
\end{itemize}

\noindent
The cut on the minimum  $P_{T \rm  ,COR}^{\rm jet}$ is dictated by the trigger. In order to avoid any possible 
bias on the measured jet shapes due to the three-level trigger selection, the thresholds
on $P_{T \rm  ,COR}^{\rm jet}$,  applied to the different data samples, have been selected such that  the 
trigger is fully efficient in the whole kinematic region under study. The measurements are performed for central 
jets in a rapidity region away from calorimeter cracks and inside the fiducial region of the CDF tracking system.

% ========================
% JET SHAPE DEFINITION
% ========================

\section{Jet shape}

\subsection{Jet shape definition}
The differential jet shape as a function of the distance $r=\sqrt{\Delta Y^2 + \Delta \phi^2}$ to the jet axis, $\rho(r)$, is defined as the average 
fraction of the jet transverse momentum that lies inside an annulus of 
inner radius $r - \delta r/2$ and outer radius $r + \delta r/2 $ 
around the jet:

\begin{equation}
\rho(r) = \frac{1}{\delta r} \frac{1}{N_{\rm jet}}\sum_{\rm jets} \frac{P_T(r - \delta r/2,r + \delta r/2) }{P_T(0,R)}, \ \ \ \ 0 \leq  r \leq R
\end{equation}
\noindent
where $N_{\rm jet}$ denotes the total number of jets, 
$P_T(r - \delta r/2,r + \delta r/2)$ is the transverse momentum within an annulus and the jet shape 
is determined for values of $r$ between 0.05 and 0.65 using $\delta r = 0.1$ intervals. The points from the differential jet shape 
at different $r$ values are correlated since, by definition, $\int_0^R \rho (r) \ \delta r = 1$.
 
The integrated jet shape, $\Psi(r)$, is defined as the average fraction of the 
jet transverse momentum that lies inside a cone of radius $r$ concentric to the jet cone:

\begin{equation}
\Psi(r) = \frac{1}{\rm N_{jet}} \sum_{\rm jets} \frac{P_T(0,r) }{P_T(0,R)}, \ \ \ \ 0 \leq  r \leq R
\end{equation}

\noindent
where, by definition, $\Psi(r = R) = 1$. 
The integrated jet shape is determined in intervals  $\delta r = 0.1$ between $r= 0$ and $r= 0.7$, and  the points at 
different $r$ values are strongly correlated.

\subsection{Jet shape reconstruction}

Calorimeter towers are used for both data and Monte Carlo simulated events 
to reconstruct the differential jet shape. For each jet, the scalar sum of 
the transverse momentum of the calorimeter towers assigned to it, $P_T(r - \delta r/2,r + \delta r/2)$, 
with a distance to the jet axis $r' = \sqrt{(Y^{\rm tower} - Y^{\rm jet})^2 + (\phi^{\rm tower}-\phi^{\rm jet})^2}$ between 
$r - \delta r/2$ and $r + \delta r/2$, is determined and divided  by 
$P_T(0, R)$. The differential jet shape, $\rho^{\rm CAL}(r)$,  
is then  determined following the prescription  in Eq.~3. Similarly, the 
integrated jet shape, $\Psi^{\rm CAL}(r)$,  is reconstructed using the calorimeter towers as defined in  Eq.~4.
The same procedure is applied to the final-state
particles in Monte Carlo generated events to reconstruct the differential and integrated 
jet shapes at the hadron level,   $\rho^{\rm HAD}_{\rm MC}(r)$ 
and $\Psi^{\rm HAD}_{\rm MC}(r)$, respectively. In the case of hadron-level jets no grid in the ($Y$-$\phi$) space has been used.

\subsection{Jet shape using charged particles}

The CDF tracking system provides an alternative method to measure the shape 
of the jets using charged particles. For each jet, tracks 
with transverse momentum, $p_{T}^{\rm track}$, above 0.5 GeV/c and pseudorapidity, $\eta^{\rm track}$,  in the 
region $|\eta^{\rm track}| < 1.4$ are assigned to it if their distances, $r$, with respect to the jet axis   
% -- $r = \sqrt{(Y^{\rm jet} - \eta^{\rm track})^2 + (\phi^{\rm jet} - \phi^{\rm track})^2}$,  
are smaller than 0.7, and  the tracks project to within 2 cm 
of the $z$-position of the primary vertex.   
The differential and integrated jet shapes,  $\rho^{\rm TRKS}(r)$ 
and $\Psi^{\rm TRKS}(r)$, are then reconstructed using the track information and following Eqs.~3~and~4. 
The measured jet shapes using tracks are employed to study systematic uncertainties on the central 
measurements as determined using calorimeter towers (see next section). Therefore,  
detailed studies have been performed on track reconstruction efficiency 
inside jets as a function of $r$ and the jet and track transverse momenta,  for both data and  
simulated events, using track embedding techniques~\cite{simon}. The difference between efficiencies in the data and Monte Carlo are  
about $3 \%$, and approximately independent of $r$ for tracks with  $0.5 \ {\rm GeV/c} < p_T^{\rm track} < 2.0 \ {\rm GeV/c}$. 
For tracks with $p_T^{\rm track} > 2.0$ GeV/c, the difference in efficiency is of the order of $5 \%$ at the core of the jet, 
decreasing as $r$ increases up to $r=0.5$. For $r>0.5$ no difference in efficiency is observed.  
The effect on the reconstructed jet shapes is smaller than $0.5 \%$ and thus has been absorbed into the systematic uncertainty.

% =======================
% UNFOLDING AND SYSTEMATIC UNCERTAINTIES
% =======================

\section{Unfolding and systematic studies}

The measured jet shapes, as determined using calorimeter towers, are corrected back 
to the hadron level using Monte Carlo samples of generated events. PYTHIA-Tune A provides a 
good description of the measured jet shapes in all regions of $P_T^{\rm jet}$ and 
is used to determine the correction factors in the unfolding procedure.  

\subsection{Jet shape corrections}

The measured jet shapes are corrected for acceptance 
and smearing effects back to the hadron level. The correction factors also account for 
the efficiency of the selection criteria and for jet reconstruction in the calorimeter.  
%Bin-by-bin correction factors are computed using PYTHIA-Tune A.
Differential and integrated 
jet shapes are reconstructed with Monte Carlo samples using both calorimeter towers,   $\rho^{\rm CAL}_{\rm MC}(r)$ 
and $\Psi^{\rm CAL}_{\rm MC}(r)$, 
and final-state particles, $\rho^{\rm HAD}_{\rm MC}(r)$ 
and $\Psi^{\rm HAD}_{\rm MC}(r)$,  in different regions of  $P_{T, \rm COR}^{\rm jet}$ and 
$P_{T, \rm HAD}^{\rm jet}$,  respectively. 
Correction factors, defined as $D(r) = \rho^{\rm HAD}_{\rm MC}(r)/\rho^{\rm CAL}_{\rm MC}(r)$ and 
$I(r) = \Psi^{\rm HAD}_{\rm MC}(r)/\Psi^{\rm CAL}_{\rm MC}(r)$, are then computed separately in 
each bin of $P_{T, \rm COR}^{\rm jet}$.
The corrected differential and integrated measurements are determined from the measured jet shapes   
as $\rho(r) =  D(r) \cdot \rho^{\rm CAL}(r)$ and  $\Psi(r) =  I(r) \cdot \Psi^{\rm CAL}(r)$.
The correction factors 
$D(r)$ do not show a significant dependence on $P_{T}^{\rm jet}$ and vary between 1.2 and 0.9 as $r$ increases.
For the integrated jet shapes, the correction factors  $I(r)$ differ from unity by less than $10 \%$ 
for $r > 0.2$.

\subsection{Systematic uncertainties}
A detailed study of the different sources of systematic uncertainties on the measured jet shapes 
has been performed~\cite{olga}: 
\begin{itemize}

\item The measured jet transverse momentum has been varied by $\pm 5 \%$ in the data to account for 
the uncertainty on the determination of the absolute energy scale in the calorimeter.  The effect on the 
measured jet shapes is of the order of $2 \%$. 

\item The unfolding procedure has been repeated using bin-by-bin correction factors extracted from 
HERWIG instead of PYTHIA-Tune A to account for any possible dependence on the modeling of parton cascades. The effect 
on the measured jet shapes is about $2 \%$ to  $5 \%$.
  
\item The ratios of uncorrected jet shape measurements as determined using calorimeter towers and tracks, 
$\rho^{\rm CAL}(r)/\rho^{\rm TRKS}(r)$ and 
$\Psi^{\rm CAL}(r)/\Psi^{\rm TRKS}(r)$, are compared
between data and Monte Carlo simulated events. 
The deviations from unity observed in the data/Monte Carlo double ratio, below $5 \%$ for the whole 
$P_T^{\rm jet}$ range, are included in the systematic uncertainties to account for the uncertainty on the description of the inactive material 
in front of the calorimeter and its response to low-energy particles.

\item The measurements are performed in different periods of Tevatron instantaneous luminosity  (between 
$0.2 \cdot  10^{31} \  {\rm cm}^{-2} {\rm s}^{-1}$ and $4 \cdot  10^{31} \  {\rm cm}^{-2} {\rm s}^{-1}$)  to 
account for possible remaining contributions from pile-up events. No significant effect is found.
\end{itemize}

\noindent
The total systematic uncertainties on $\rho (r)$ and $\Psi(r)$ have been computed for the different $r$ ranges 
by adding in quadrature the deviations from the central values. The statistical uncertainties are negligible compared to the systematic errors except for jets with $P_{T}^{\rm jet} > 300$ GeV/c. The systematic uncertainties have been added in quadrature to 
the statistical errors and the total uncertainties are shown in the figures. The total uncertainty in the measured 
data points, for different $P_{T}^{\rm jet}$ and $r$ ranges, varies between $5 \%$ to $10 \%$ except for jets with $P_{T}^{\rm jet} > 300$ GeV/c for which the total error is above $20 \%$.  
% ====================== 
% RESULTS
% ======================
\section{Results}
The corrected differential and integrated jet shapes,  $\rho (r)$ 
and $\Psi (r)$, refer to midpoint jets at the hadron level with 
cone size $R=0.7$ in the region $0.1 < |Y^{\rm jet}| < 0.7$ and 
$37$~GeV/c~$< P_T^{\rm jet} < 380$~GeV/c.

\subsection{Comparison with Monte Carlo}

\begin{figure}[tbh]
\centerline{\includegraphics{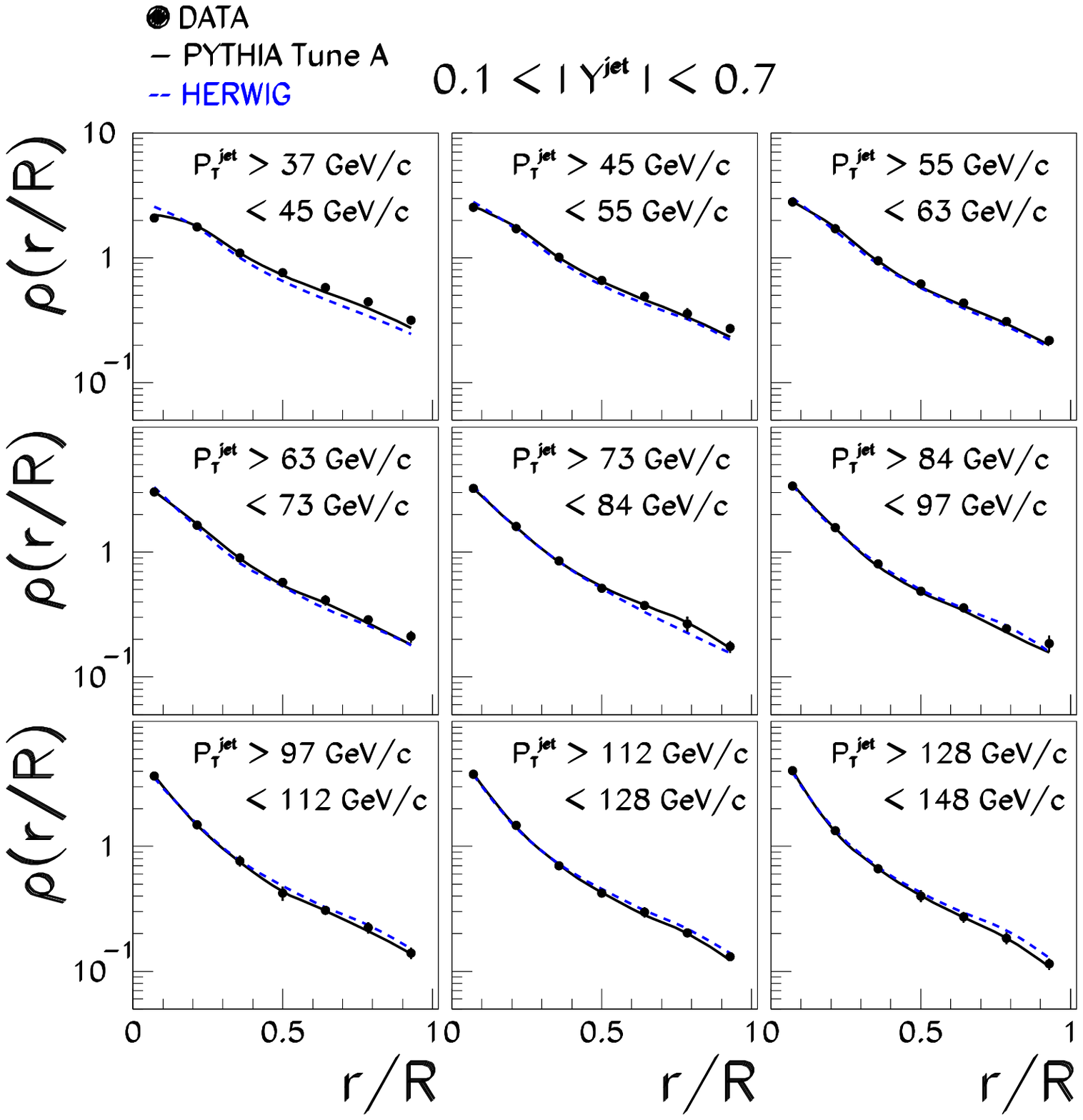}} 

\vspace{-1 cm}
\caption{The measured differential jet shape, $\rho(r/R)$, in inclusive jet production for jets 
with $0.1 < |Y^{\rm jet}| < 0.7$ and $37 \ {\rm GeV/c} < P_T^{\rm jet} < 148  \ {\rm GeV/c}$,  
is shown in different $P_T^{\rm jet}$ regions. Error bars indicate the statistical and systematic uncertainties added in quadrature.
The predictions of PYTHIA-Tune A (solid lines) and HERWIG (dashed lines) are shown for comparison.} 
\label{fig2}
\end{figure}

\begin{figure}[tbh]
\centerline{\includegraphics{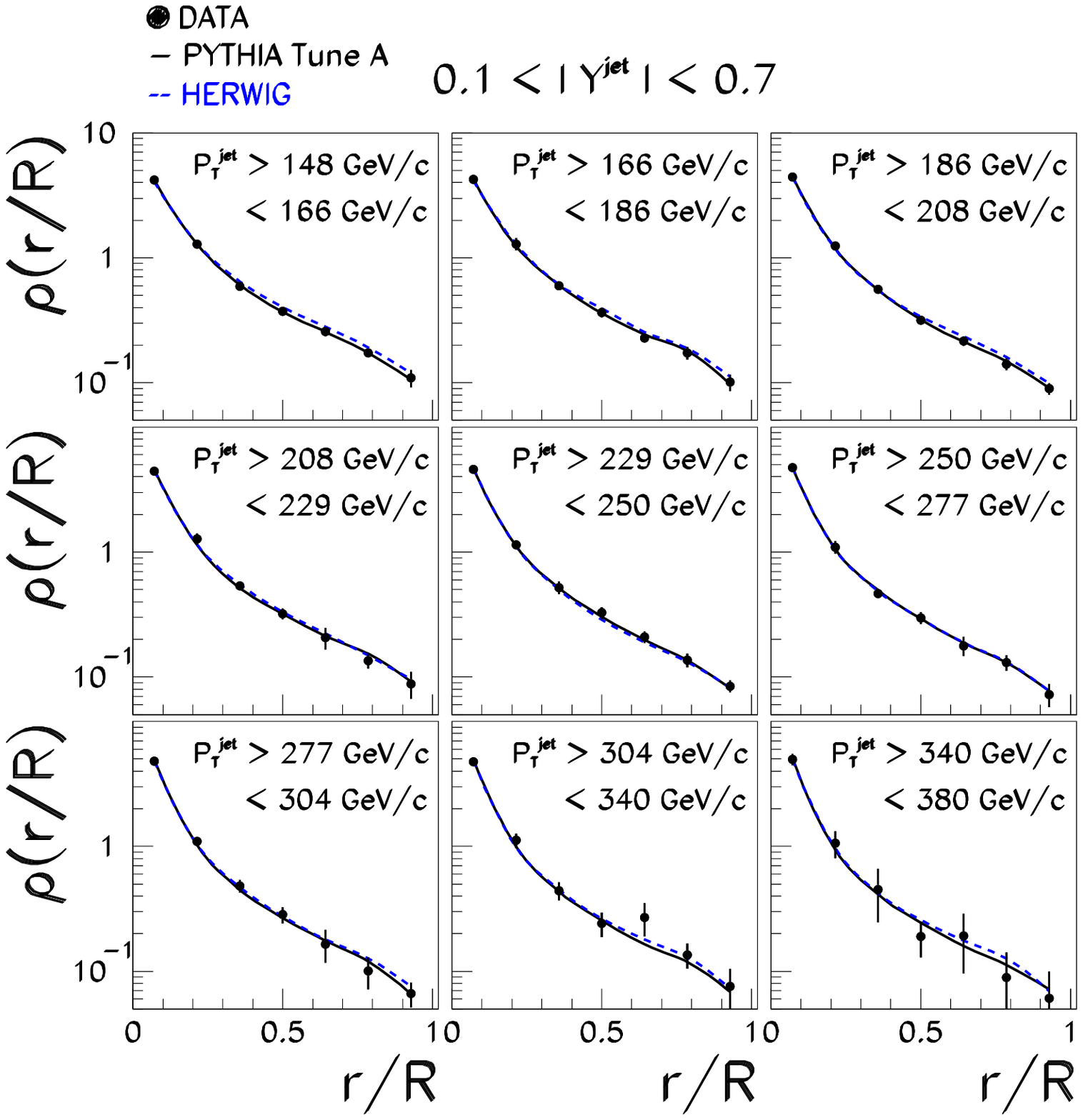}} 

\vspace{-1.5 cm}
\caption{The measured differential jet shape, $\rho(r/R)$, in inclusive jet production for jets 
with $0.1 < |Y^{\rm jet}| < 0.7$ and $148 \ {\rm GeV/c} < P_T^{\rm jet} < 380 \ {\rm GeV/c}$,  
is shown in different $P_T^{\rm jet}$ regions. Error bars indicate the statistical and systematic uncertainties added in quadrature.
The predictions of PYTHIA-Tune A (solid lines) and HERWIG (dashed lines) are shown for comparison.} 
\label{fig3}
\end{figure}

Figures~\ref{fig2} and \ref{fig3} show the measured differential jet shapes, $\rho(r/R)$,  in  bins  
of $P_{T}^{\rm jet}$ for jets in the region 
$0.1 < |Y^{\rm jet}| < 0.7$ and $37 \ {\rm GeV/c}< P_{T}^{\rm jet} < 380 \ {\rm GeV/c}$,  compared to the PYTHIA-Tune A and 
HERWIG Monte Carlo  
predictions at the hadron level. The measured jet shapes show a prominent peak at low $r$ which indicates that the 
majority of the jet momentum is concentrated at $r/R < 0.2$. At low  $P_{T}^{\rm jet}$,  the fraction of transverse
momentum at  the core of the jet is about a factor of 6 times larger than that at the tail. This factor increases at higher 
$P_{T}^{\rm jet}$  and  is of the order of 100 for jets with  $P_{T}^{\rm jet} > 340 \ {\rm GeV/c}$. 
PYTHIA-Tune A provides a good description of the measured jet shapes in all regions of $P_{T}^{\rm jet}$. 
The jets predicted by HERWIG follow the  measurements but tend to be narrower 
than the data at low $P_{T}^{\rm jet}$. The latter can be attributed to the absence of additional soft contributions from 
multiple parton interactions in HERWIG, which  are particularly important at low $P_{T}^{\rm jet}$.  

\begin{figure}[tbh]
\centerline{\includegraphics{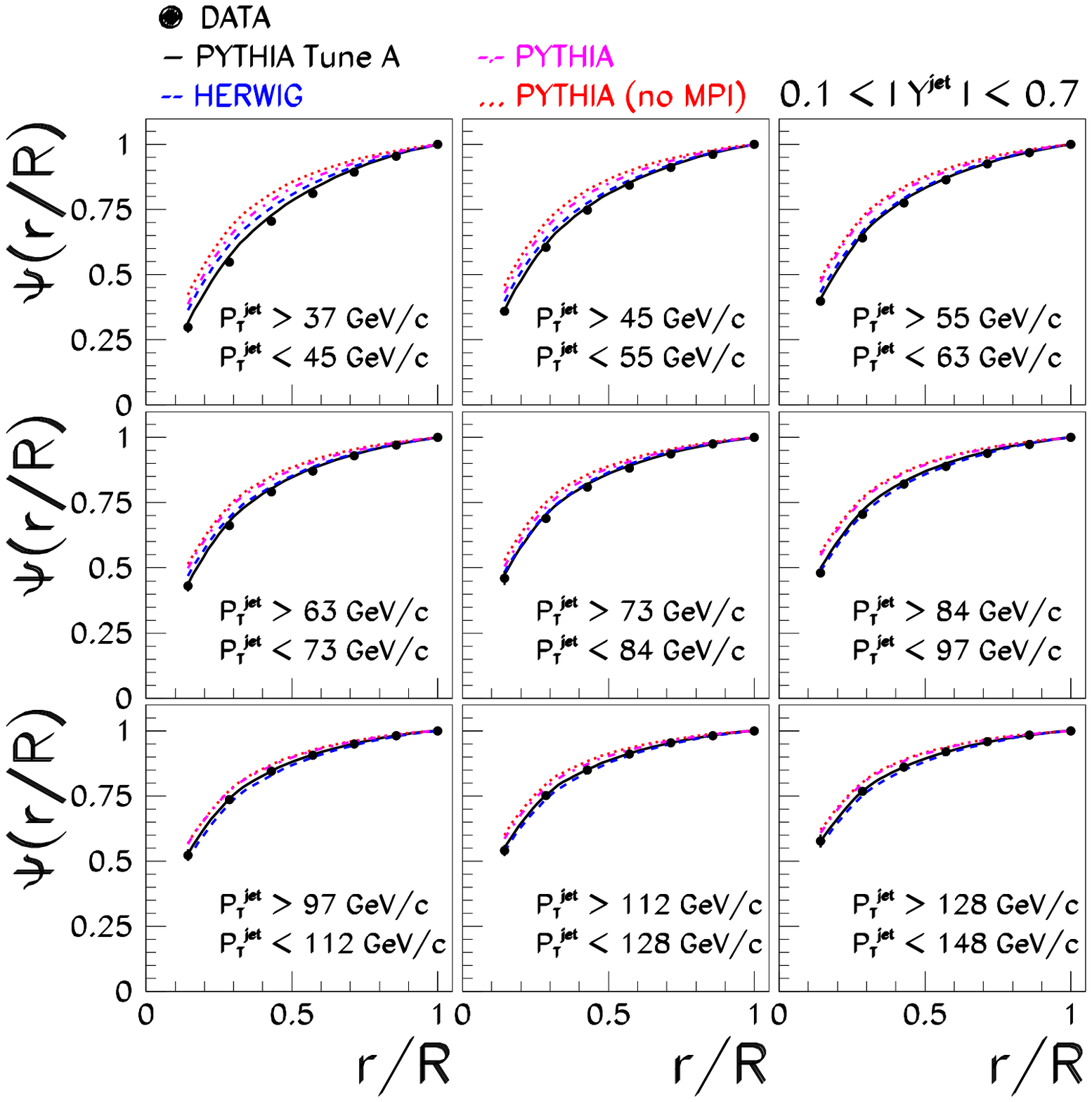}} 

\vspace{-1.5 cm}
\caption{The measured integrated jet shape, $\Psi(r/R)$, in inclusive jet production for jets 
with $0.1 < |Y^{\rm jet}| < 0.7$ and $37 \ {\rm GeV/c} < P_T^{\rm jet} < 148 \ {\rm GeV/c}$,  
is shown in different $P_T^{\rm jet}$ regions.  Error bars indicate the statistical and 
systematic uncertainties added in quadrature. The predictions of PYTHIA-Tune A (solid lines), PYTHIA (dashed-dotted lines), 
PYTHIA-(no MPI) (dotted lines) and HERWIG (dashed lines) 
are shown for comparison.} 
\label{fig4}
\end{figure}

\begin{figure}[tbh]
\centerline{\includegraphics{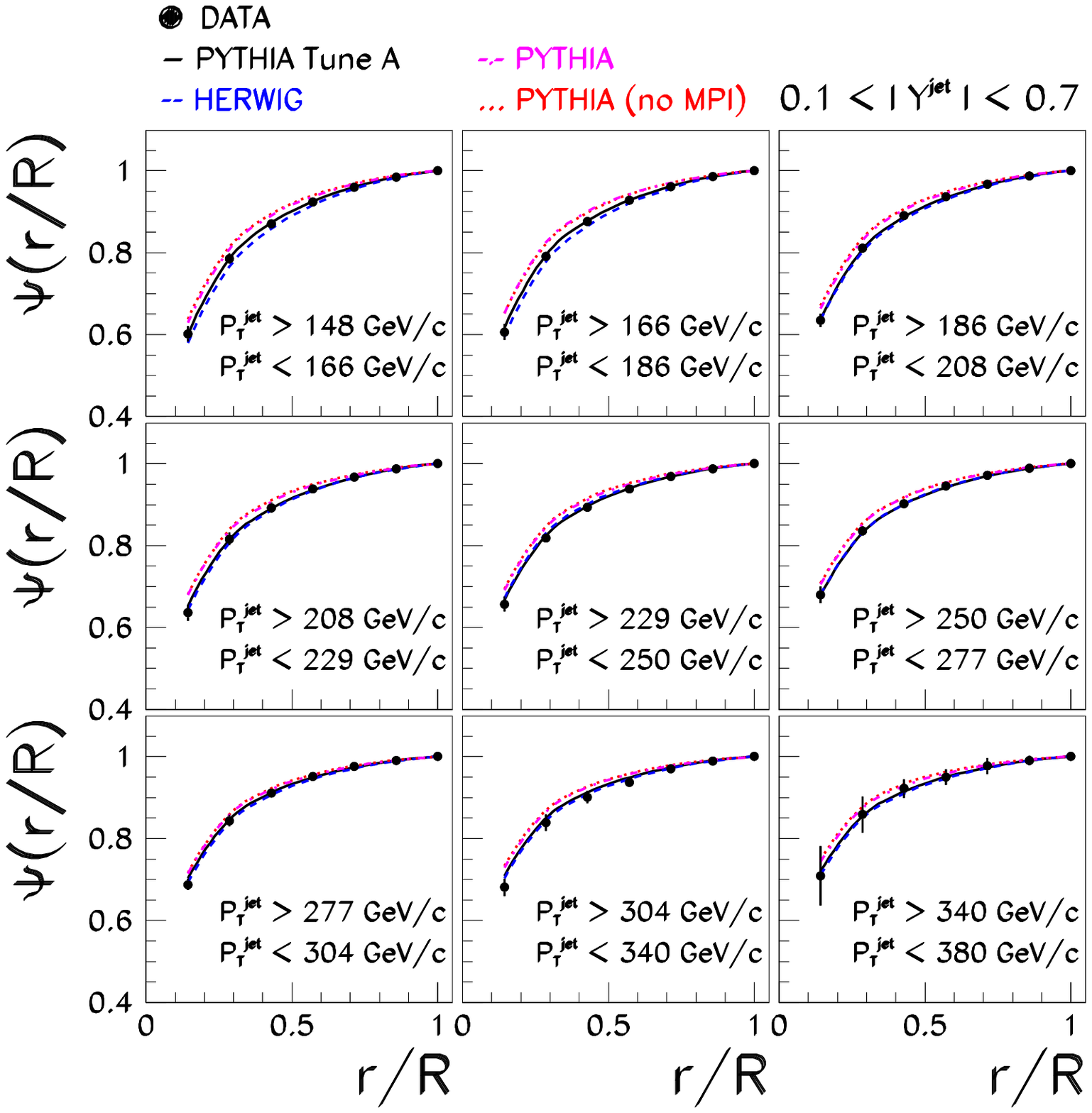}} 

\vspace{-1.5 cm}
\caption{The measured integrated jet shape, $\Psi(r/R)$, in inclusive jet production for jets 
with $0.1 < |Y^{\rm jet}| < 0.7$ and $148 \ {\rm GeV/c} < P_T^{\rm jet} < 380 \ {\rm GeV/c}$,  
is shown in different $P_T^{\rm jet}$ regions.  Error bars indicate the statistical and 
systematic uncertainties added in quadrature. The predictions of PYTHIA-Tune A (solid lines), PYTHIA (dashed-dotted lines), 
PYTHIA-(no MPI) (dotted lines) and HERWIG (dashed lines) 
are shown for comparison.} 
\label{fig5}
\end{figure}

Figures~\ref{fig4} and \ref{fig5} present the measured integrated jet shapes, $\Psi(r/R)$,  in bins  
of $P_{T}^{\rm jet}$, for jets with $0.1 < |Y^{\rm jet}| < 0.7$ and  
$37 \ {\rm GeV/c}< P_{T}^{\rm jet} < 380 \ {\rm GeV/c}$, compared to 
HERWIG, PYTHIA-Tune A,  PYTHIA and PYTHIA-(no MPI) predictions, to  illustrate the importance of a proper modeling of soft-gluon 
radiation in describing the measured jet shapes. 

\begin{figure}[tbh]
\centerline{\includegraphics{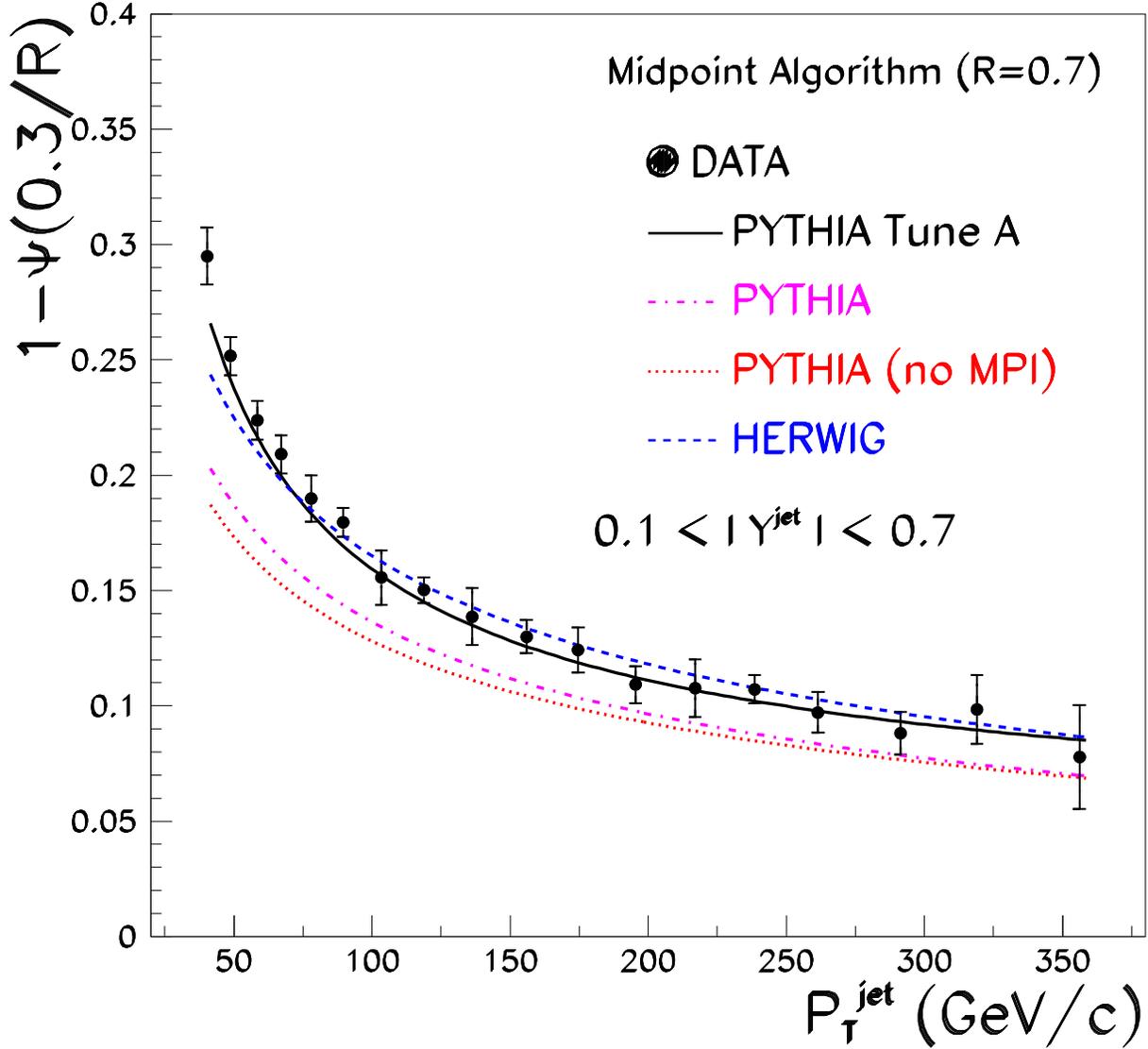}} 

\vspace{-1.5 cm}
\caption{The measured $1 - \Psi(0.3/R)$ as a function of $P_T^{\rm jet}$ 
for jets with $0.1 < |Y^{\rm jet}| < 0.7$ and $37 \ {\rm GeV/c} < P_T^{\rm jet} < 380 \ {\rm GeV/c}$.
Error bars indicate the statistical and systematic uncertainties added in quadrature. 
The predictions of PYTHIA-Tune A (solid line) 
, PYTHIA (dashed-dotted line), 
PYTHIA-(no MPI) (dotted line) and HERWIG (dashed line) 
are shown for comparison.} 
\label{fig6}
\end{figure}

Figure~\ref{fig6} shows, for a fixed radius $r_0 = 0.3$, the average 
fraction of the jet transverse momentum outside $r=r_0$, ($1-\Psi(r_0/R)$), as a function of   
$P_{T}^{\rm jet}$. The points are located at the weighted mean in each $P_{T}^{\rm jet}$ range.
The measurements show that the fraction of jet transverse momentum inside a given fixed $r_0/R$ increases ($1-\Psi(r_0/R)$ decreases) 
with $P_{T}^{\rm jet}$, indicating  
that the jets become narrower as  $P_{T}^{\rm jet}$ increases. PYTHIA with default parameters 
produces jets systematically narrower than the data in the whole region in $P_{T}^{\rm jet}$. The contribution from  
secondary parton interactions between remnants to the predicted jet shapes (as shown by the difference between  PYTHIA and PYTHIA-(no MPI) predictions) is important at low $P_{T}^{\rm jet}$. PYTHIA-Tune A predictions describe all of the data well 
(a $\chi^2$ test in Fig.~\ref{fig6} gives a value of 13.6 for a total of 18 data points). HERWIG describes 
the measured jet shapes well but produces jets slightly narrower than the data at low  $P_{T}^{\rm jet}$. This results in a 
significantly higher $\chi^2$ value of 33.8 for 18 data points.

\subsection{Quark- and gluon-jet contributions}

\begin{figure}[tbh]
\centerline{\includegraphics{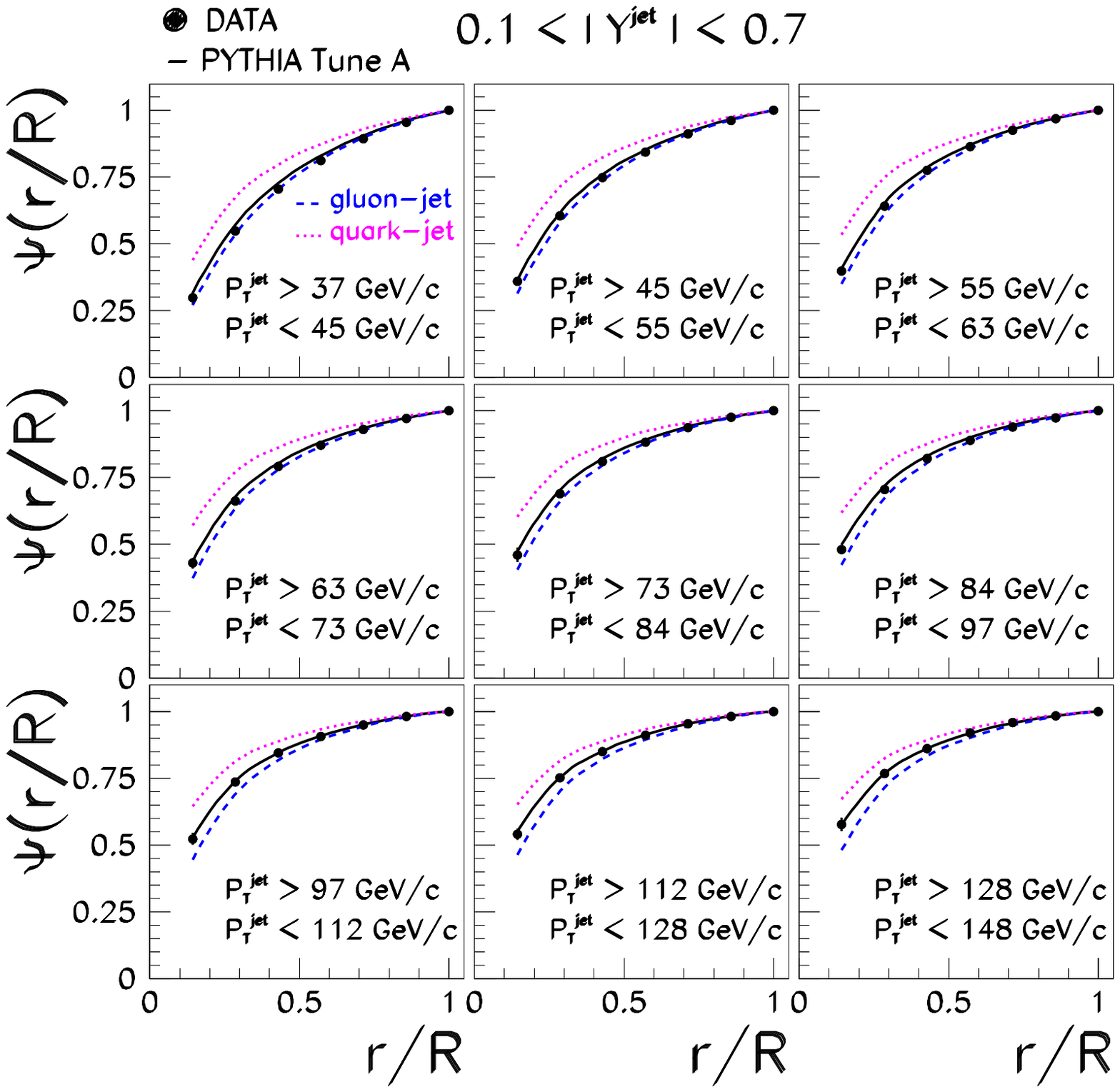}} 

\vspace{-1.5 cm}
\caption{The measured integrated jet shape, $\Psi(r/R)$, in inclusive jet production for jets 
with $0.1 < |Y^{\rm jet}| < 0.7$ and $37 \ {\rm GeV/c} < P_T^{\rm jet} < 148 \ {\rm GeV/c}$,  
is shown in different $P_T^{\rm jet}$ regions.  Error bars indicate the statistical and 
systematic uncertainties added in quadrature. The predictions of PYTHIA-Tune A (solid lines) and the 
separate predictions for quark-initiated jets (dotted lines) and gluon-initiated jets (dashed lines)
are shown for comparison.} 
\label{fig7}
\end{figure}

\begin{figure}[tbh]
\centerline{\includegraphics{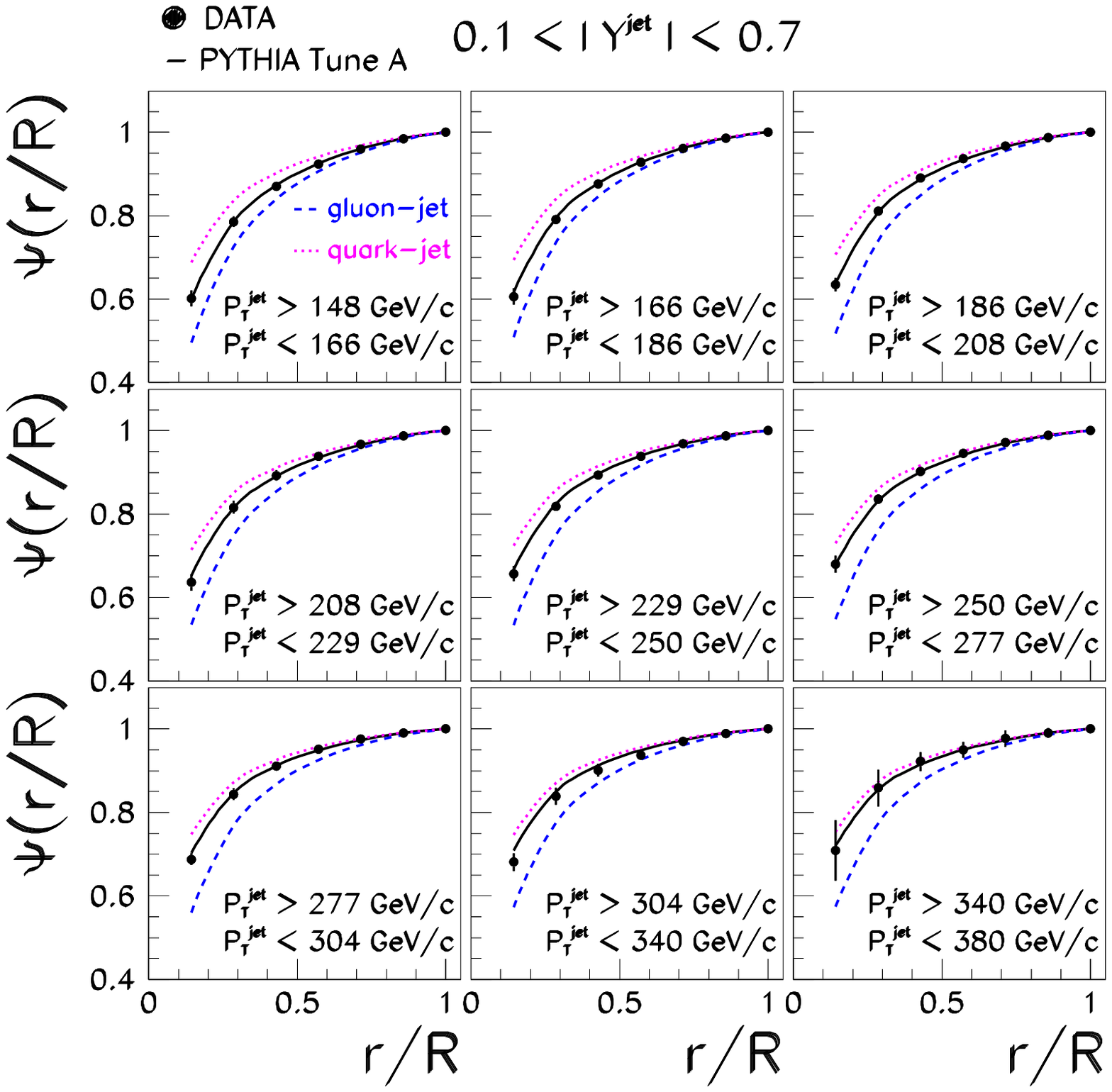}} 

\vspace{-1.5 cm}
\caption{The measured integrated jet shape, $\Psi(r/R)$, in inclusive jet production for jets 
with $0.1 < |Y^{\rm jet}| < 0.7$ and $148 \ {\rm GeV/c} < P_T^{\rm jet} < 380 \ {\rm GeV/c}$,  
is shown in different $P_T^{\rm jet}$ regions.  Error bars indicate the statistical and 
systematic uncertainties added in quadrature. The predictions of PYTHIA-Tune A (solid lines) and the 
separate predictions for quark-initiated jets (dotted lines) and gluon-initiated jets (dashed lines)
are shown for comparison.} 
\label{fig8}
\end{figure}

\begin{figure}[tbh]
\centerline{\includegraphics{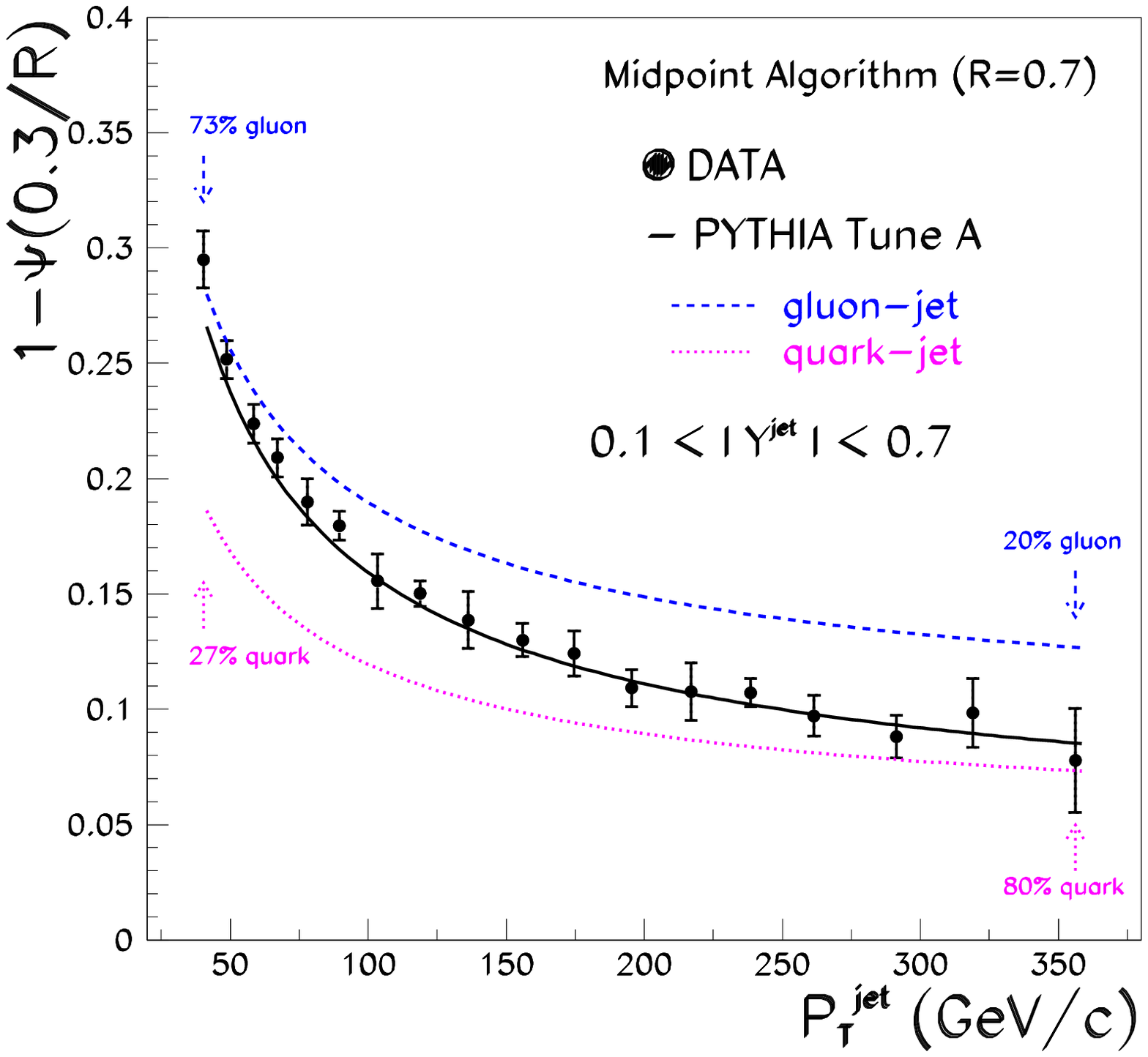}} 

\vspace{-1.5 cm}
\caption{The measured  $1 - \Psi(0.3/R)$ as a function of $P_T^{\rm jet}$ 
for jets with $0.1 < |Y^{\rm jet}| < 0.7$ and $37 \ {\rm GeV/c} < P_T^{\rm jet} < 380 \ {\rm GeV/c}$.
Error bars indicate the statistical and systematic uncertainties added in quadrature. 
The predictions of PYTHIA-Tune A (solid line) and the 
separate predictions for quark-initiated jets (dotted line) and gluon-initiated jets (dashed line)
are shown for comparison. The arrows indicate the fraction of quark- and gluon-initiated jets at low and very high $P_T^{\rm jet}$, as predicted by PYTHIA-Tune A.} 
\label{fig9}
\end{figure}

Figures~\ref{fig7} and \ref{fig8} present the measured integrated jet shapes, $\Psi(r/R)$, in bins
of $P_{T}^{\rm jet}$, for jets with $0.1 < |Y^{\rm jet}| < 0.7$ and  $37 \ {\rm GeV/c}< P_{T}^{\rm jet} < 380 \ {\rm GeV/c}$, compared to 
PYTHIA-Tune A predictions (as in Figs. \ref{fig4} and \ref{fig5}). In these figures, predictions are also shown separately for 
quark- and gluon-jets.
Each hadron-level jet from PYTHIA is classified as a quark- or gluon-jet by matching ($Y$-$\phi$ plane) its 
direction with that of one of the outgoing partons from 
the hard interaction. 
The Monte Carlo predictions indicate that, for  the jets used in this analysis, the measured jet shapes are dominated by 
contributions from gluon-initiated jets at low $P_{T}^{\rm jet}$ while contributions from   
quark-initiated jets become important at high $P_{T}^{\rm jet}$. This can be explained in terms of 
the different partonic contents in the proton and antiproton  contributing to the low- and high-$P_{T}^{\rm jet}$ regions, since the 
mixture of gluon- and quark-jet in the final state  partially reflects 
the nature of the incoming partons that participate in the hard interaction. 
Figure~\ref{fig9} shows the measured $1-\Psi(r_0/R)$, $r_0 = 0.3$, as a function of   
$P_{T}^{\rm jet}$ compared to  PYTHIA-Tune A predictions with  
quark- and gluon-jets shown separately. The trend with  $P_{T}^{\rm jet}$ in the measured jet shapes   
is mainly attributed to the different quark- and gluon-jet mixture in the final state and  perturbative QCD 
effects related to the running of the strong coupling, $\alpha_s(P_{T}^{\rm jet})$~\cite{ee}. The Monte Carlo 
predicts that the fraction of gluon-initiated jets decreases from about 73~$\%$ at low $P_{T}^{\rm jet}$ to
20~$\%$ at very high $P_{T}^{\rm jet}$, while the fraction of quark-initiated jets increases.

% =======================
% SUMMARY AND CONCLUSIONS
% =======================
\section{Summary and conclusions}
Jet shapes have been measured in inclusive jet production in $p \overline{p}$ collisions for jets 
in the kinematic region~$37 \  {\rm GeV/c} < P_{T}^{\rm jet} < 380 \ {\rm GeV/c}$ and $0.1 < |Y^{\rm jet}| < 0.7$. 
Jets become narrower as $P_{T}^{\rm jet}$ increases  
which can be mainly attributed to the change in the quark- and
gluon-jet mixture in the final state and the running
of the strong coupling with $P_T^{\rm jet}$.
PYTHIA Monte Carlo predictions, using default parameters, do not give a good description of 
the measured jet shapes in the entire $P_{T}^{\rm jet}$ range.
PYTHIA-Tune A, which includes  enhanced contributions from initial-state 
gluon radiation and  secondary parton interactions between remnants, describes the data better. 
HERWIG gives a reasonable description of the measured jet shapes but tends to  
produce jets that are too narrow at low $P_{T}^{\rm jet}$ which can be attributed to the absence 
of soft contributions from multiple parton interactions in HERWIG. Jet shape measurements thus can be used to introduce strong 
constraints on phenomenological models  describing soft-gluon radiation and the underlying event 
in hadron-hadron interactions.

% ======================
% Acknowledgments
% ======================

\section*{Acknowledgments}

We thank the Fermilab staff and the technical staffs of the participating institutions for their vital contributions. This work was supported by the U.S. Department of Energy and National Science Foundation; the Italian Istituto Nazionale di Fisica Nucleare; the Ministry of Education, Culture, Sports, Science and Technology of Japan; the Natural Sciences and Engineering Research Council of Canada; the National Science Council of the Republic of China; the Swiss National Science Foundation; the A.P. Sloan Foundation; the Bundesministerium fuer Bildung und Forschung, Germany; the Korean Science and Engineering Foundation and the Korean Research Foundation; the Particle Physics and Astronomy Research Council and the Royal Society, UK; the Russian Foundation for Basic Research; the Comision Interministerial de Ciencia y Tecnologia, Spain; in part by the European Community's Human Potential Programme under contract HPRN-CT-2002-00292; and the Academy of Finland.

% =======================
% BIBLIOGRAPHY
% =======================

\clearpage

% ========================
% FIGURES
% ========================

\clearpage

\clearpage

\clearpage

\clearpage

\clearpage

\clearpage

\clearpage

\clearpage

\clearpage


\begin{thebibliography}{99.}
\bibitem{rapidity} The pseudorapidity is defined as $\eta = - {\rm ln(tan(\frac{\theta}{2})) }$, where the polar
angle $\theta$ is taken with respect to the proton beam direction. The rapidity is defined as $Y = \rm \frac{1}{2} ln(\frac{E+p_z}{E-p_z}) $, where 
$E$ denotes the energy and $p_z$ is the component of the momentum along the proton beam direction.
\bibitem{hadron}  The hadronic final state in the Monte Carlo generators is defined 
using particles with lifetime above $10^{-11} {\rm s}$.
\end{thebibliography}
\end{document}